%% file: Theta5.tex
\documentclass[letterpaper,11pt]{article}
\usepackage{fullpage}
\usepackage[english]{babel}
\usepackage{graphicx}
\usepackage{amssymb}
\usepackage{amsmath}
\usepackage{xspace}
\usepackage{wrapfig}
\usepackage{hyperref}

\graphicspath{{figures/}}

\newcommand{\etal}{\emph{et al.}\xspace}
\newcommand{\graph}{\ensuremath{\theta_5}-graph\xspace}
\newcommand{\T}[2]{\ensuremath{T_{#1 #2}}}

\newcommand{\valc}{\ensuremath{2 \left( 2 + \sqrt{5} \right) \approx 8.472}\xspace}
\newcommand{\valsr}{\ensuremath{\sqrt{50 + 22 \sqrt{5}} \approx 9.960}\xspace}
\newcommand{\vallb}{\ensuremath{\frac{1}{2}(11\sqrt{5} -17) \approx 3.798}\xspace}

\newtheorem{defin}{Definition}
  
\newtheorem{theo}[defin]{Theorem}
  \newenvironment{theorem}{\begin{theo} \sl}{\end{theo}}
\newtheorem{lem}[defin]{Lemma}
  \newenvironment{lemma}{\begin{lem} \sl}{\end{lem}}
\newtheorem{coro}[defin]{Corollary}
  
\newtheorem{obs}[defin]{Observation}
  
\newenvironment{myproof}{\emph{Proof.}}{\hfill $\Box$ \medskip\\}

\author{Prosenjit Bose\thanks{School of Computer Science, Carleton University. Email: \texttt{jit@scs.carleton.ca, morin@scs.carleton.ca, andre@cg.scs.carleton.ca, sander@cg.scs.carleton.ca}. Research supported in part by NSERC.}
\and
\addtocounter{footnote}{-1}
Pat Morin\footnotemark
\and
\addtocounter{footnote}{-1}
Andr\'e van Renssen\footnotemark
\and
\addtocounter{footnote}{-1}
Sander Verdonschot\footnotemark
}
\title{The \graph is a spanner}
\date{\today}

\begin{document}

\maketitle

\begin{abstract}
 Given a set of points in the plane, we show that the $\theta$-graph with 5 cones is a geometric spanner with spanning ratio at most \valsr. This is the first constant upper bound on the spanning ratio of this graph. The upper bound uses a constructive argument that gives a (possibly self-intersecting) path between any two vertices, of length at most $\sqrt{50 + 22 \sqrt{5}}$ times the Euclidean distance between the vertices. We also give a lower bound on the spanning ratio of \vallb.
\end{abstract}

\section{Introduction}
\label{sec:Introduction}

\begin{wrapfigure}{r}{40mm}
  \vspace{-2em}
  \begin{center}
    \includegraphics{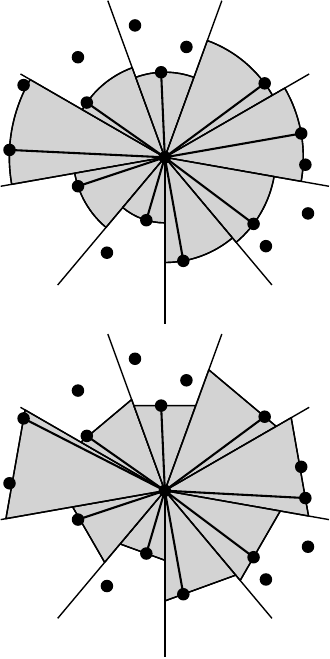}
  \end{center}
  \vspace{-1em}
  \caption{(Top) The construction of the Yao graph. (Bottom) The construction of the $\theta$-graph.}
  \vspace{-1em}
  \label{fig:cones}
\end{wrapfigure}

A \emph{$t$-spanner} of a weighted graph $G$ is a spanning subgraph $H$ with the property that for all pairs of vertices, the weight of the shortest path between the vertices in $H$ is at most $t$ times the weight of the shortest path in $G$. The \emph{spanning ratio} of $H$ is the smallest $t$ for which it is a $t$-spanner. We say that a graph is a \emph{spanner} if it has a finite spanning ratio. The graph $G$ is referred to as the \emph{underlying graph}. In this paper, the underlying graph is the complete graph on a finite set of $n$ points in the plane and the weight of an edge is the Euclidean distance between its endpoints. A spanner of such a graph is called a \emph{geometric spanner}. For a comprehensive overview of geometric spanners, we refer the reader to the book by Narasimhan and Smid \cite{NS06}.

One simple way to build a geometric spanner is to first partition the plane around each vertex into a fixed number of cones and then add an edge between the vertex and the closest vertex in each cone (see Figure~\ref{fig:cones}, top). The resulting graph is called a Yao graph, and is typically denoted by $Y_k$, where $k$ is the number of cones around each vertex. If the number of cones is sufficiently large, we can find a path between any two vertices by starting at one and walking to the closest vertex in the cone that contains the other, then repeating this until we reach the destination. Intuitively, this results in a short path because we are always walking approximately in the right direction, and, since our neighbour is the closest vertex in that direction, never too far. However, when the number of cones is small, the path found in this way can be very long (see Section~\ref{sec:lbrouting}).

Yao graphs were introduced independently by Flinchbaugh and Jones~\cite{flinchbaugh1981strong} and Yao~\cite{yao1982constructing}, before the concept of spanners was even introduced by Chew in 1986~\cite{chew1986there}. To the best of our knowledge, the first proof that Yao graphs are geometric spanners was published in 1993, by Alth{\"o}fer~\etal~\cite{althofer1993sparse}. In particular, they showed that for every $t > 1$, there exists a $k$ such that $Y_k$ is a $t$-spanner. It appears that some form of this result was known earlier, as Clarkson~\cite{clarkson1987approximation} already remarked in 1987 that $Y_{12}$ is a $1 + \sqrt{3}$-spanner, albeit without providing a proof or reference. In 2004, Bose~\etal~\cite{bose2004approximating} provided a more specific bound on the spanning ratio, by showing that for $k > 8$, $Y_k$ is a geometric spanner with spanning ratio at most $1 / (\cos \theta - \sin \theta)$, where $\theta = 2\pi/k$. This bound was later improved to $1 / (1 - 2 \sin(\theta / 2))$, for $k > 6$~\cite{bose2012piArxiv}.

If we modify the definition of Yao graphs slightly, by connecting not to the closest point in each cone, but to the point whose projection on the bisector of that cone is closest (see Figure~\ref{fig:cones}, bottom), we obtain another type of geometric spanner, called a $\theta$-graph. These graphs were introduced independently by Clarkson~\cite{clarkson1987approximation} and Keil~\cite{keil1988approximating,keil1992classes}, who preferred them to Yao graphs because they are easier to compute. It turns out that $\theta$-graphs share most of the properties of Yao graphs: there is a constant $k$ for every $t > 1$ such that $\theta_k$ is a $t$-spanner, and the spanning ratio for $k > 6$ is $1 / (1 - 2 \sin(\theta / 2))$~\cite{ruppert1991approximating}. Very recently, the bounds on $\theta$-graphs were even pushed beyond those on Yao graphs~\cite{bose2013spanning}, including a matching upper- and lower bound of $1 + 2 \sin(\theta / 2)$ for all $\theta_k$-graphs with $k \geq 6$ and $k \equiv 2 \pmod 4$~\cite{bose2012optimal}.

Although most early research focused on Yao and $\theta$-graphs with a large number of cones, using the smallest possible number of cones is important for many practical applications, where the cost of a network is mostly determined by the number of edges. One such example is point-to-point wireless networks. These networks use narrow directional wireless transceivers that can transmit over long distances (up to 50km \cite{dist1,dist2}). The cost of an edge in such a network is therefore equal to the cost of the two transceivers that are used at each endpoint of that edge. In such networks, the cost of building a $\theta_6$-graph is approximately 29\% higher than the cost of building a \graph if the transceivers are randomly distributed~\cite{morin2013average}. This leads to the natural question: for which values of $k$ are $Y_k$ and $\theta_k$ spanners? Kanj~\cite{kanj2013geometric} presented this question as one of the main open problems in the area of geometric spanners.

\begin{wrapfigure}{r}{40mm}
  \vspace{-2em}
  \begin{center}
    \includegraphics{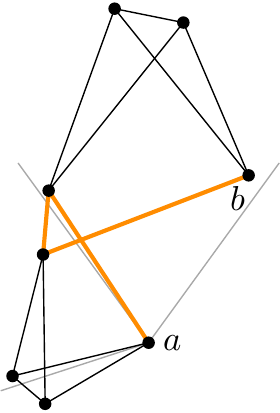}
  \end{center}
  \vspace{-1em}
  \caption{A $\theta_5$-graph where the shortest path between two vertices (in bold) crosses itself.}
  \vspace{-1em}
  \label{fig:cross}
\end{wrapfigure}

Surprisingly, this question was not studied until quite recently. In 2009, El~Molla~\cite{el2009yao} showed that, for $k < 4$, there is no constant $t$ such that $Y_k$ is a $t$-spanner. These proofs translate to $\theta$-graphs as well. Bonichon~\etal~\cite{bonichon2010connections} showed that $\theta_6$ is a 2-spanner, and this is tight. This result was later used by Damian and Raudonis~\cite{damian2012yao} to show that $Y_6$ is a spanner as well. Bose~\etal~\cite{bose2012pi} showed that $Y_4$ is a spanner, and a flurry of recent activity has led to the same result for both $\theta_4$~\cite{barba2013stretch} and $Y_5$~\cite{barba2014new}.

In this paper we present the final piece of this puzzle, by giving the first constant upper bound on the spanning ratio of the \graph, thereby proving that it is a geometric spanner. Since the proof is constructive, it gives us a path between any two vertices, $u$ and $w$, of length at most $\valsr \cdot |uw|$. Surprisingly, this path can cross itself, a property we observed for the shortest path as well (see Figure~\ref{fig:cross}).

After completion of this research, we discovered that some form of this result appears to have been known already in 1991, as Ruppert and Seidel~\cite{ruppert1991approximating} mention that they could prove a bound near 10 on the spanning ratio of $\theta_5$. However, their paper does not include a proof and, to the best of our knowledge, they have not published one since.

In addition to the upper bound on the spanning ratio, we prove two lower bounds. We give an example of a point set where the \graph has a spanning ratio of \vallb, and we show that the traditional $\theta$-routing algorithm (follow the edge to the closest vertex in the cone that contains the destination) can result in very long paths, even though a short path exists.

\section{Connectivity}
\label{sec:Connectivity}

In this section we first give a more precise definition of the \graph, before proving that it is connected.

\begin{figure}[ht]
  \centering
  \includegraphics{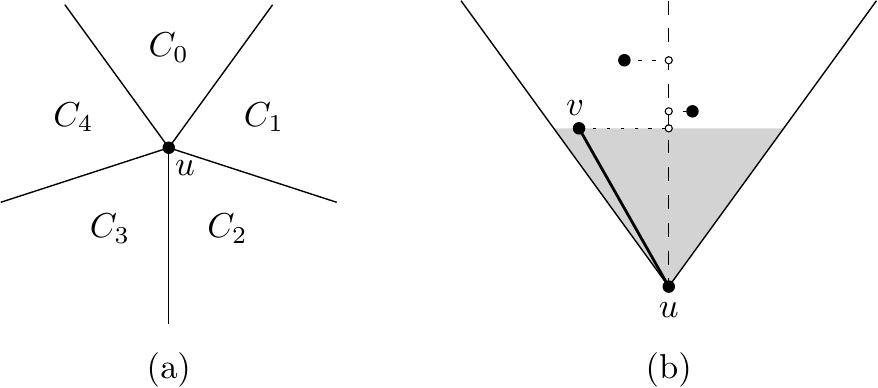}
  \caption{(a) The cones of a vertex $u$. (b) The vertex $v$ is closest to $u$. The shaded region is the canonical triangle $\T{u}{v}$.}
  \label{fig:theta5cones}
\end{figure}

Given a set $P$ of points in the plane, we consider each point $u \in P$ and partition the plane into 5 cones (regions in the plane between two rays originating from the same point) with apex $u$, each defined by two rays at consecutive multiples of $\theta=2\pi/5$ radians from the negative $y$-axis. We label the cones $C_0$ through $C_4$, in clockwise order around $u$, starting from the top (see~Figure~\ref{fig:theta5cones}a). If the apex is not clear from the context, we use $C_i^u$ to denote cone $C_i$ with apex $u$. For the sake of brevity, we typically write ``a cone \emph{of} $u$'' instead of ``a cone with apex $u$''.

The \graph is then built by considering each vertex $u$ and connecting it with an edge to the `closest' vertex in each of its cones, where distance is measured by projecting each vertex onto the bisector of that cone (see Figure~\ref{fig:theta5cones}b). We use this definition of \emph{closest} in the remainder of the paper.

For simplicity, we assume that no two points lie on a line parallel or perpendicular to a cone boundary. This guarantees that each vertex connects to at most one vertex in each cone, and thus that the graph has at most $5n$ edges. For any set of points that does not satisfy this assumption, there exists a tiny angle such that the assumption holds if we rotate all cones by this angle. In terms of the graph, this rotation is equivalent to a tie-breaking rule that always selects the candidate that comes last in clockwise order. Thus, our conclusions about the spanning ratio hold in either case, even though our proofs rely on the general position assumption.

Given two vertices $u$ and $v$, we define their \emph{canonical triangle} $\T{u}{v}$ to be the triangle bounded by the cone of $u$ that contains $v$ and the line through $v$ perpendicular to the bisector of that cone. For example, the shaded region in Figure~\ref{fig:theta5cones}b is the canonical triangle $\T{u}{v}$. Note that for any pair of vertices $u$ and $v$, there are two canonical triangles: $\T{u}{v}$ and $\T{v}{u}$. We equate the size $|\T{u}{v}|$ of a canonical triangle to the length of one of the sides incident to the apex $u$. This gives us the useful property that any line segment between $u$ and a point inside the triangle has length at most $|\T{u}{v}|$.

To introduce the structure of the main proof, we first show that the \graph is connected.

\begin{theorem}
 \label{thm:connected}
 The \graph is connected.
\end{theorem}
\begin{myproof}
 We prove that there is a path between any (ordered) pair of vertices in the \graph, using induction on the size of their canonical triangle. Formally, given two vertices $u$ and $w$, we perform induction on the rank (relative position) of $\T{u}{w}$ among the canonical triangles of all pairs of vertices, when ordered by size. For ease of description, we assume that $w$ lies in the right half of $C_0^u$. The other cases are analogous.

 If $\T{u}{w}$ has rank 1, it is the smallest canonical triangle. Therefore there can be no point closer to $u$ in $C_0^u$, so the \graph must contain the edge $(u, w)$. This proves the base case.

 If $\T{u}{w}$ has a larger rank, our inductive hypothesis is that there exists a path between any pair of vertices with a smaller canonical triangle. Let $a$ and $b$ be the left and right corners of $\T{u}{w}$. Let $m$ be the midpoint of $ab$ and let $x$ be the intersection of $ab$ and the bisector of $\angle mub$ (see~Figure~\ref{fig:canon}a).

 \begin{figure}[ht]
  \centering
  \includegraphics{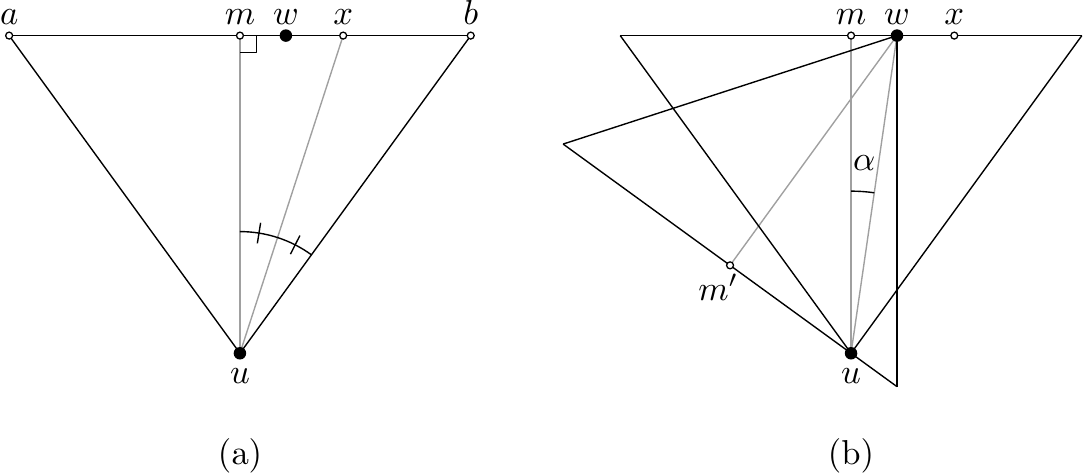}
  \caption{(a) The canonical triangle $\T{u}{w}$. (b) If $w$ lies between $m$ and $x$, $\T{w}{u}$ is smaller than $\T{u}{w}$.}
  \label{fig:canon}
 \end{figure}

 If $w$ lies to the left of $x$, consider the canonical triangle $\T{w}{u}$. Let $m'$ be the midpoint of the side of $\T{w}{u}$ opposite $w$ and let $\alpha = \angle muw$ (see~Figure~\ref{fig:canon}b). Note that $\angle uwm' = \frac{\pi}{5} - \alpha$, since $um$ and the vertical border of $\T{w}{u}$ are parallel and both are intersected by $uw$. Using basic trigonometry, we can express the size of $\T{w}{u}$ as follows.
 \[ |\T{w}{u}|
~~=~~\frac{|wm'|}{\cos \frac{\pi}{5}}
~~=~~\frac{\cos \angle uwm' \cdot |uw|}{\cos \frac{\pi}{5}}
~~=~~\frac{\cos \left( \frac{\pi}{5} - \alpha \right) \cdot \frac{|um|}{\cos \alpha}}{\cos \frac{\pi}{5}}
~~=~~\frac{\cos \left( \frac{\pi}{5} - \alpha \right)}{\cos \alpha} \cdot |\T{u}{w}|
 \]
 Since $w$ lies to the left of $x$, the angle $\alpha$ is less than $\pi/10$, which means that $\cos ( \frac{\pi}{5} - \alpha ) / \cos \alpha$ is less than 1. Hence $\T{w}{u}$ is smaller than $\T{u}{w}$ and by induction, there is a path between $w$ and $u$. Since the \graph is undirected, we are done in this case. The rest of the proof deals with the case where $w$ lies on or to the right of $x$.

 If $\T{w}{u}$ is empty, there is an edge between $u$ and $w$ and we are done, so assume that this is not the case. Then there is a vertex $v_w$ that is closest to $w$ in $C_3^w$ (the cone of $w$ that contains $u$). This gives rise to four cases, depending on the location of $v_w$ (see Figure~\ref{fig:cases}a). In each case, we will show that $\T{u}{v_w}$ is smaller than $\T{u}{w}$ and hence we can apply induction to obtain a path between $u$ and $v_w$. Since $v_w$ is the closest vertex to $w$ in $C_3$, there is an edge between $v_w$ and $w$, completing the path between $u$ and $w$.

 \begin{figure}[ht]
  \centering
  \includegraphics{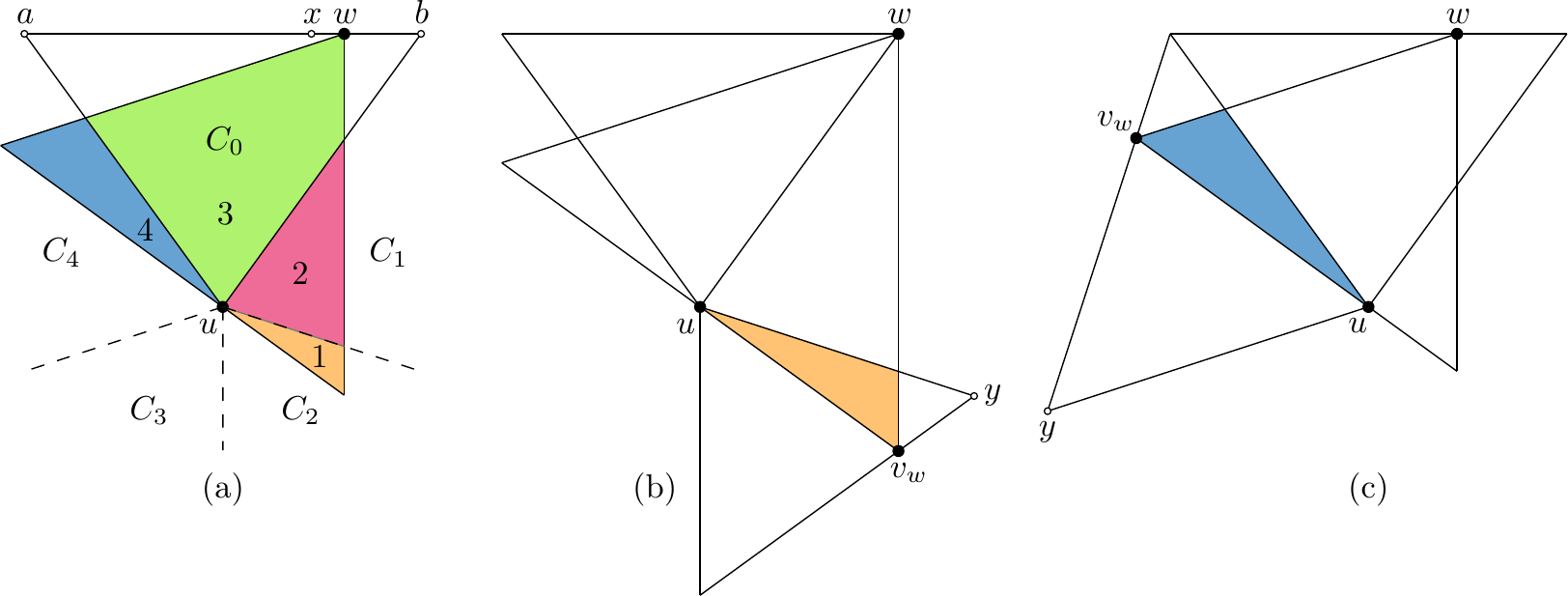}
  \caption{(a) The four cases for $v_w$. (b) Case 1: The situation that maximizes $|\T{u}{v_w}|$ when $v_w$ lies in $C_2^u$. (c) Case 4: The situation that maximizes $|\T{u}{v_w}|$ when $v_w$ lies in $C_4^u$.}
  \label{fig:cases}
 \end{figure}

 \paragraph{Case 1.} $v_w$ lies in $C_2^u$. In this case, the size of $\T{u}{v_w}$ is maximized when $v_w$ lies in the bottom right corner of $\T{w}{u}$ and $w$ lies on $b$. Let $y$ be the rightmost corner of $\T{u}{v_w}$ (see~Figure~\ref{fig:cases}b). Using the law of sines, we can express the size of $\T{u}{v_w}$ as follows.
 \[
  |\T{u}{v_w}|
~~=~~|uy|
~~=~~\frac{\sin \angle uv_wy}{\sin \angle uyv_w} \cdot |uv_w|
~~=~~\frac{\sin \frac{3\pi}{5}}{\sin \frac{3\pi}{10}} \cdot \tan \frac{\pi}{5} \cdot |\T{u}{w}|
~~<~~|\T{u}{w}|
 \]
 \paragraph{Case 2.} $v_w$ lies in $C_1^u$. In this case, the size of $\T{u}{v_w}$ is maximized when $w$ lies on $b$ and $v_w$ lies almost on $w$. By symmetry, this gives $|\T{u}{v_w}| = |\T{u}{w}|$. However, $v_w$ cannot lie precisely on $w$ and must therefore lie a little closer to $u$, giving us that $|\T{u}{v_w}| < |\T{u}{w}|$.
 \paragraph{Case 3.} $v_w$ lies in $C_0^u$. As in the previous case, the size of $\T{u}{v_w}$ is maximized when $v_w$ lies almost on $w$, but since $v_w$ must lie closer to $u$, we have that $|\T{u}{v_w}| < |\T{u}{w}|$.
 \paragraph{Case 4.} $v_w$ lies in $C_4^u$. In this case, the size of $\T{u}{v_w}$ is maximized when $v_w$ lies in the left corner of $\T{w}{u}$ and $w$ lies on $x$. Let $y$ be the bottom corner of $\T{u}{v_w}$ (see~Figure~\ref{fig:cases}c). Since $x$ is the point where $|\T{u}{w}| = |\T{w}{u}|$, and $v_wyuw$ forms a parallelogram, $|\T{u}{v_w}| = |\T{u}{w}|$. However, by general position, $v_w$ cannot lie on the boundary of $\T{w}{u}$, so it must lie a little closer to $u$, giving us that $|\T{u}{v_w}| < |\T{u}{w}|$.
 
 \bigskip
 Since any vertex in $C_3^u$ would be further from $w$ than $u$ itself, these four cases are exhaustive.
\end{myproof}

\section{Spanning ratio}
\label{sec:SpanningRatio}

In this section, we prove an upper bound on the spanning ratio of the \graph.

\begin{lemma}
 \label{lem:spanningPath}
 Between any pair of vertices $u$ and $w$ of a \graph, there is a path of length at most $c \cdot |\T{u}{w}|$, where $c = \valc$.
\end{lemma}
\begin{myproof}
 We begin in a way similar to the proof of Theorem~\ref{thm:connected}. Given an ordered pair of vertices $u$ and $w$, we perform induction on the size of their canonical triangle. If $|\T{u}{w}|$ is minimal, there must be a direct edge between them. Since $c > 1$ and any edge inside $\T{u}{w}$ with endpoint $u$ has length at most $|\T{u}{w}|$, this proves the base case. The rest of the proof deals with the inductive step, where we assume that there exists a path of length at most $c \cdot |T|$ between every pair of vertices whose canonical triangle $T$ is smaller than $\T{u}{w}$. As in the proof of Theorem~\ref{thm:connected}, we assume that $w$ lies in the right half of $C_0^u$. If $w$ lies to the left of $x$, we have seen that $\T{w}{u}$ is smaller than $\T{u}{w}$. Therefore we can apply induction to obtain a path of length at most $c \cdot |\T{w}{u}| < c \cdot |\T{u}{w}|$ between $u$ and $w$. Hence we need to concern ourselves only with the case where $w$ lies on or to the right of $x$.

 If $u$ is the vertex closest to $w$ in $C_3^w$ or $w$ is the closest vertex to $u$ in $C_0^u$, there is a direct edge between them and we are done by the same reasoning as in the base case. Therefore assume that this is not the case and let $v_w$ be the vertex closest to $w$ in $C_3^w$. We distinguish the same four cases for the location of $v_w$ (see~Figure~\ref{fig:cases}a). We already showed that we can apply induction on $\T{u}{v_w}$ in each case. This is a crucial part of the proof for the first three cases.

 Most of the cases come down to finding a path between $u$ and $w$ of length at most $(g + h \cdot c) \cdot |\T{u}{w}|$, for constants $g$ and $h$ with $h < 1$. The smallest value of $c$ for which this is bounded by $c \cdot |\T{u}{w}|$ is $g/(1-h)$. If this is at most \valc, we are done.

 \paragraph{Case 1.} $v_w$ lies in $C_2^u$. By induction, there exists a path between $u$ and $v_w$ of length at most $c \cdot |\T{u}{v_w}|$. Since $v_w$ is the closest vertex to $w$ in $C_3^w$, there is a direct edge between them, giving a path between $u$ and $w$ of length at most $|wv_w| + c \cdot |\T{u}{v_w}|$.
 
 Given any initial position of $v_w$ in $C_2^u$, we can increase $|wv_w|$ by moving $w$ to the right. Since this does not change $|\T{u}{v_w}|$, the worst case occurs when $w$ lies on $b$. Then we can increase both $|wv_w|$ and $|\T{u}{v_w}|$ by moving $v_w$ into the bottom corner of $\T{w}{u}$. This gives rise to the same worst-case configuration as in the proof of Theorem~\ref{thm:connected}, depicted in Figure~\ref{fig:cases}b. Building on the analysis there, we can bound the worst-case length of the path as follows.
 \[
     |wv_w| + c \cdot |\T{u}{v_w}|
~~=~~\frac{|\T{u}{w}|}{\cos \frac{\pi}{5}} + c \cdot \frac{\sin \frac{3\pi}{5}}{\sin \frac{3\pi}{10}} \cdot \tan \frac{\pi}{5} \cdot |\T{u}{w}|
 \]
 This is at most $c \cdot |\T{u}{w}|$ for $c \geq 2 \left( 2 + \sqrt{5} \right)$. Since we picked $c = 2 \left( 2 + \sqrt{5} \right)$,  the theorem holds in this case. Note that this is one of the cases that determines the value of $c$.
 
 \begin{figure}[ht]
  \centering
  \includegraphics{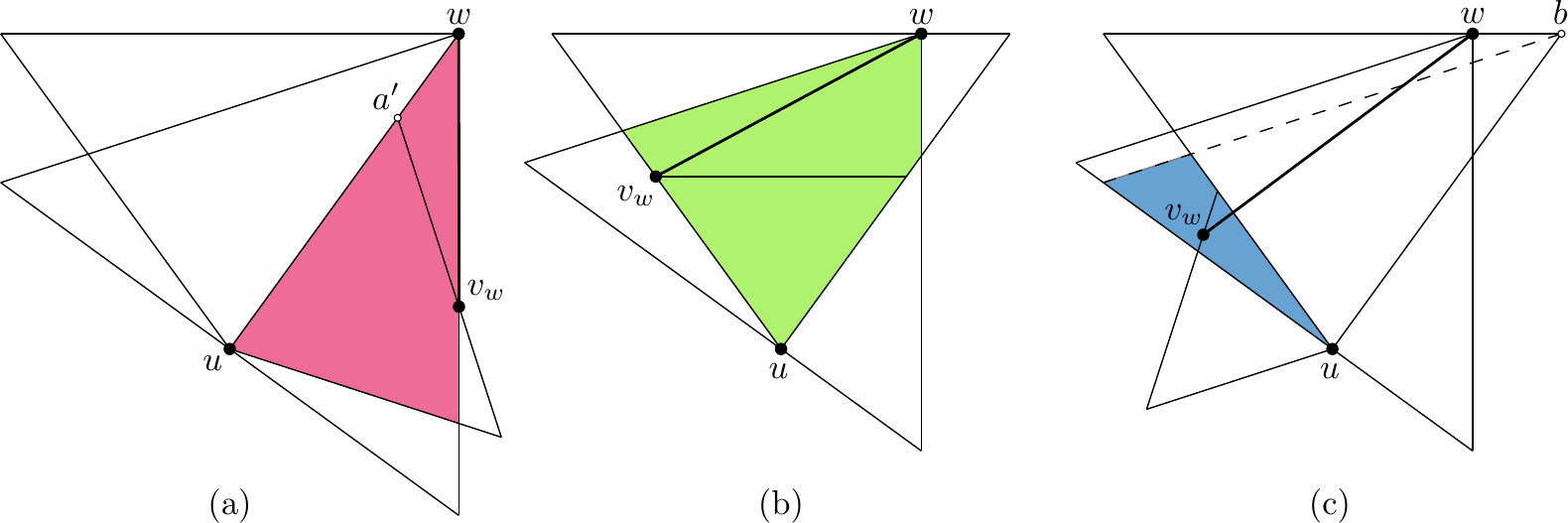}
  \caption{(a) Case 2: Vertex $v_w$ lies on the boundary of $C_3^w$ after moving it down along the side of $\T{u}{v_w}$. (b) Case 3: Vertex $v_w$ lies on the boundary of $C_0^u$ after moving it left along the side of $\T{u}{v_w}$. (c) Case 4: Vertex $v_w$ lies in $C_4^u \cap C_3^b$.}
  \label{fig:worst}
 \end{figure}
 
 \paragraph{Case 2.} $v_w$ lies in $C_1^u$. By the same reasoning as in the previous case, we have a path of length at most $|wv_w| + c \cdot |\T{u}{v_w}|$ between $u$ and $w$ and we need to bound this length by $c \cdot |\T{u}{w}|$.
 
 Given any initial position of $v_w$ in $C_1^u$, we can increase $|wv_w|$ by moving $w$ to the right. Since this does not change $|\T{u}{v_w}|$, the worst case occurs when $w$ lies on $b$. We can further increase $|wv_w|$ by moving $v_w$ down along the side of $\T{u}{v_w}$ opposite $u$ until it hits the boundary of $C_1^u$ or $C_3^w$, whichever comes first (see~Figure~\ref{fig:worst}a).
 
 Now consider what happens when we move $v_w$ along these boundaries. If $v_w$ lies on the boundary of $C_1^u$ and we move it away from $u$ by $\Delta$, $|\T{u}{v_w}|$ increases by $\Delta$. At the same time, $|wv_w|$ might decrease, but not by more than $\Delta$. Since $c > 1$, the total path length is maximized by moving $v_w$ as far from $u$ as possible, until it hits the boundary of $C_3^w$. Once $v_w$ lies on the boundary of $C_3^w$, we can express the size of $\T{u}{v_w}$ as follows, where $a'$ is the top corner of $\T{u}{v_w}$.
 \[
  |\T{u}{v_w}|
  ~~=~~ |\T{u}{w}| - |wa'|
  ~~=~~ |\T{u}{w}| - |wv_w| \cdot \frac{\sin \angle wv_wa'}{\sin \angle wa'v_w}
  ~~=~~ |\T{u}{w}| - |wv_w| \cdot \frac{\sin \frac{\pi}{10}}{\sin \frac{7\pi}{10}}
 \]
Now we can express the length of the complete path as follows.
 \[
  |wv_w| + c \cdot |\T{u}{v_w}|
  ~~=~~ |wv_w| + c \cdot \left( |\T{u}{w}| - |wv_w| \cdot \frac{\sin \frac{\pi}{10}}{\sin \frac{7\pi}{10}} \right)
  ~~=~~ c \cdot |\T{u}{w}| - \left( c \cdot \frac{\sin \frac{\pi}{10}}{\sin \frac{7\pi}{10}} - 1 \right) \cdot |wv_w|
 \]
Since $c > \sin \frac{7\pi}{10} / \sin \frac{\pi}{10} \approx 2.618$, we have that $c \cdot (\sin \frac{\pi}{10} / \sin \frac{7\pi}{10}) - 1 > 0$. Therefore $|wv_w| + c \cdot |\T{u}{v_w}| < c \cdot |\T{u}{w}|$.
 
 \paragraph{Case 3.} $v_w$ lies in $C_0^u$. Again, we have a path of length at most $|wv_w| + c \cdot |\T{u}{v_w}|$ between $u$ and $w$ and we need to bound this length by $c \cdot |\T{u}{w}|$.
 
 Given any initial position of $v_w$ in $C_0^u$, moving $v_w$ to the left increases $|wv_w|$ while leaving $|\T{u}{v_w}|$ unchanged. Therefore the path length is maximized when $v_w$ lies on the boundary of either $C_0^u$ or $C_3^w$, whichever it hits first (see~Figure~\ref{fig:worst}b).
 
 Again, consider what happens when we move $v_w$ along these boundaries. Similar to the previous case, if $v_w$ lies on the boundary of $C_0^u$ and we move it away from $u$ by $\Delta$, $|\T{u}{v_w}|$ increases by $\Delta$, while $|wv_w|$ might decrease by at most $\Delta$. Since $c > 1$, the total path length is maximized by moving $v_w$ as far from $u$ as possible, until it hits the boundary of $C_3^w$. Once there, the situation is symmetric to the previous case, with $|\T{u}{v_w}| = |\T{u}{w}| - |wv_w| \cdot (\sin \frac{\pi}{10} / \sin \frac{7\pi}{10})$. Therefore the theorem holds in this case as well.
 
 \paragraph{Case 4.} $v_w$ lies in $C_4^u$. This is the hardest case. Similar to the previous two cases, the size of $\T{u}{v_w}$ can be arbitrarily close to that of $\T{u}{w}$, but in this case $|wv_w|$ does not approach $0$. This means that simply invoking the inductive hypothesis on $\T{u}{v_w}$ does not work, so another strategy is required. We first look at a subcase where we \emph{can} apply induction directly, before considering the position of $v_u$, the closest vertex to $u$ in $C_0$.
 
 \paragraph{Case 4a.} $v_w$ lies in $C_4^u \cap C_3^b$. This situation is illustrated in Figure~\ref{fig:worst}c. Given any initial position of $v_w$, moving $w$ to the right onto $b$ increases the total path length by increasing $|wv_w|$ while not affecting $|\T{u}{v_w}|$. Here we use the fact that $v_w$ already lies in $C_3^b$, otherwise we would not be able to move $w$ onto $b$ while keeping $v_w$ in $C_3^w$. Now the total path length is maximized by placing $v_w$ on the left corner of $\T{w}{u}$. Since this situation is symmetrical to the worst-case situation in Case 1, the theorem holds by the same analysis.
 
 \bigskip
 Next, we distinguish four cases for the position of $v_u$ (the closest vertex to $u$ in $C_0$), illustrated in Figure~\ref{fig:closestcases}a. The cases are: (4b) $w$ lies in $C_4^{v_u}$, (4c) $w$ lies in $C_0^{v_u}$, (4d) $w$ lies in $C_1^{v_u}$ and $v_u$ lies in $C_3^w$, and (4e) $w$ lies in $C_1^{v_u}$ and $v_u$ lies in $C_4^w$. These are exhaustive, since the cones $C_4$, $C_0$ and $C_1$ are the only ones that can contain a vertex above the current vertex, and $w$ must lie above $v_u$, as $v_u$ is closer to $u$. Further, if $w$ lies in $C_1^{v_u}$, $v_u$ must lie in one of the two opposite cones of $w$.
 
 We can solve the first two cases by applying our inductive hypothesis to $\T{v_u}{w}$.
 
 \begin{figure}[ht]
  \centering
  \includegraphics{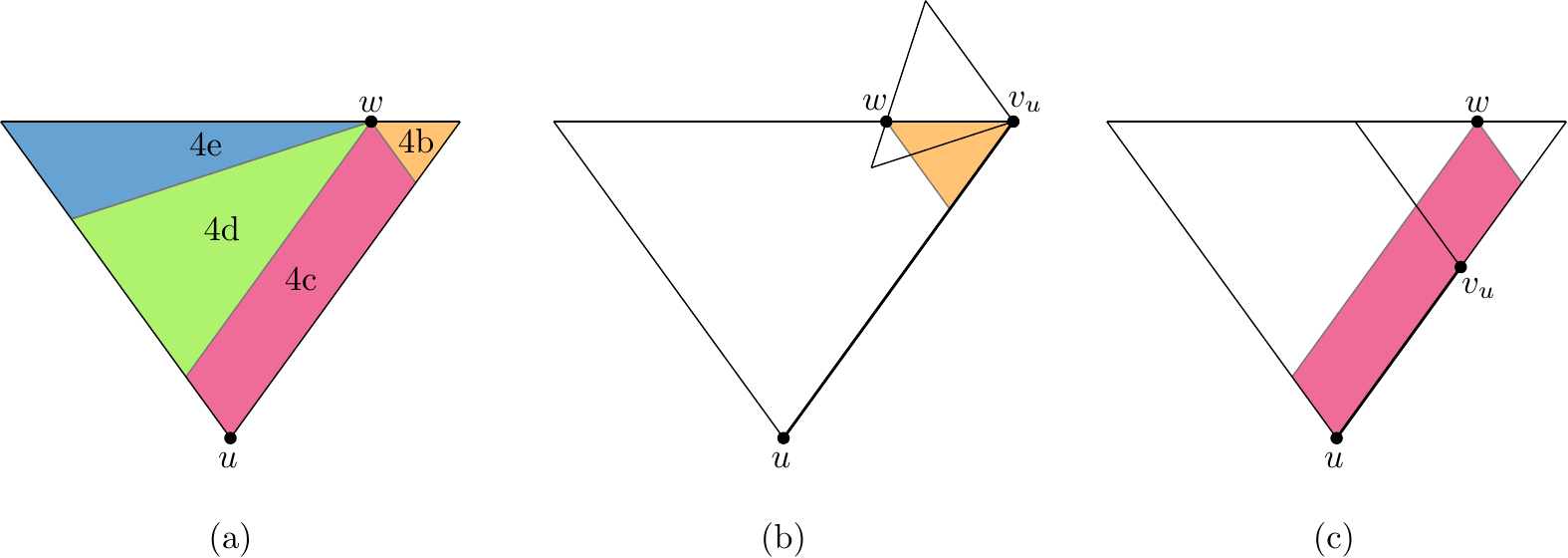}
  \caption{(a) The four different cases for the position of $v_u$. (b) The worst-case configuration with $w$ in $C_4^{v_u}$. (c) A configuration with $w$ in $C_0^{v_u}$, after moving $v_u$ onto the right side of $C_0^u$.}
  \label{fig:closestcases}
 \end{figure}
 
 \paragraph{Case 4b.} $w$ lies in $C_4^{v_u}$. To apply our inductive hypothesis, we need to show that $|\T{v_u}{w}| < |\T{u}{w}|$. If that is the case, we obtain a path between $v_u$ and $w$ of length at most $c \cdot |\T{v_u}{w}|$. Since $v_u$ is the closest vertex to $u$, there is a direct edge from $u$ to $v_u$, resulting in a path between $u$ and $w$ of length at most $|uv_u| + c \cdot |\T{v_u}{w}|$.
 
 Given any intial positions for $v_u$ and $w$, moving $w$ to the left increases $|\T{v_u}{w}|$ while leaving $|uv_u|$ unchanged. Moving $v_u$ closer to $b$ increases both. Therefore the path length is maximal when $w$ lies on $x$ and $v_u$ lies on $b$ (see~Figure~\ref{fig:closestcases}b). Using the law of sines, we can express $|\T{v_u}{w}|$ as follows.
 \[
  |\T{v_u}{w}|
  ~~=~~ \frac{\sin \frac{3\pi}{5}}{\sin \frac{3\pi}{10}} \cdot |wv_u|
  ~~=~~ \frac{\sin \frac{3\pi}{5}}{\sin \frac{3\pi}{10}} \cdot \frac{\sin \frac{\pi}{10}}{\sin \frac{3\pi}{5}} \cdot |\T{u}{w}|
  ~~=~~ \frac{\sin \frac{\pi}{10}}{\sin \frac{3\pi}{10}} \cdot |\T{u}{w}|
  ~~=~~ \frac{1}{2} \left( 3 - \sqrt{5} \right) \cdot |\T{u}{w}|
 \]
 Since $\frac{1}{2} \left( 3 - \sqrt{5} \right) < 1$, we have that $|\T{v_u}{w}| < |\T{u}{w}|$ and we can apply our inductive hypothesis to $\T{v_u}{w}$. Since $|uv_u| = |\T{u}{w}|$, the complete path has length at most $c \cdot |\T{u}{w}|$ for
 \[
  c
  ~~\geq~~ \frac{1}{1 - \frac{1}{2} \left( 3 - \sqrt{5} \right)}
  ~~=~~ \frac{1}{2} \left( 1 + \sqrt{5} \right)
  ~~\approx~~ 1.618.
 \]
 
 \paragraph{Case 4c.} $w$ lies in $C_0^{v_u}$. Since $v_u$ lies in $C_0^u$, it is clear that $|\T{v_u}{w}| < |\T{u}{w}|$, which allows us to apply our inductive hypothesis. This gives us a path between $u$ and $w$ of length at most $|uv_u| + c \cdot |\T{v_u}{w}|$. For any initial location of $v_u$, we can increase the total path length by moving $v_u$ to the right until it hits the side of $C_0^u$ (see~Figure~\ref{fig:closestcases}c), since $|\T{v_u}{w}|$ stays the same and $|uv_u|$ increases. Once there, we have that $|uv_u| + |\T{v_u}{w}| = |\T{u}{w}|$. Since $c > 1$, this immediately implies that $|uv_u| + c \cdot |\T{v_u}{w}| \leq c \cdot |\T{u}{w}|$, proving the theorem for this case.

 \bigskip
 To solve the last two cases, we need to consider the positions of both $v_u$ and $v_w$. Recall that for $v_w$, there is only a small region left where we have not yet proved the existance of a short path between $u$ and $w$. In particular, this is the case when $v_w$ lies in cone $C_4^u$, but not in $C_3^b$.

 \begin{figure}[ht]
  \centering
  \includegraphics{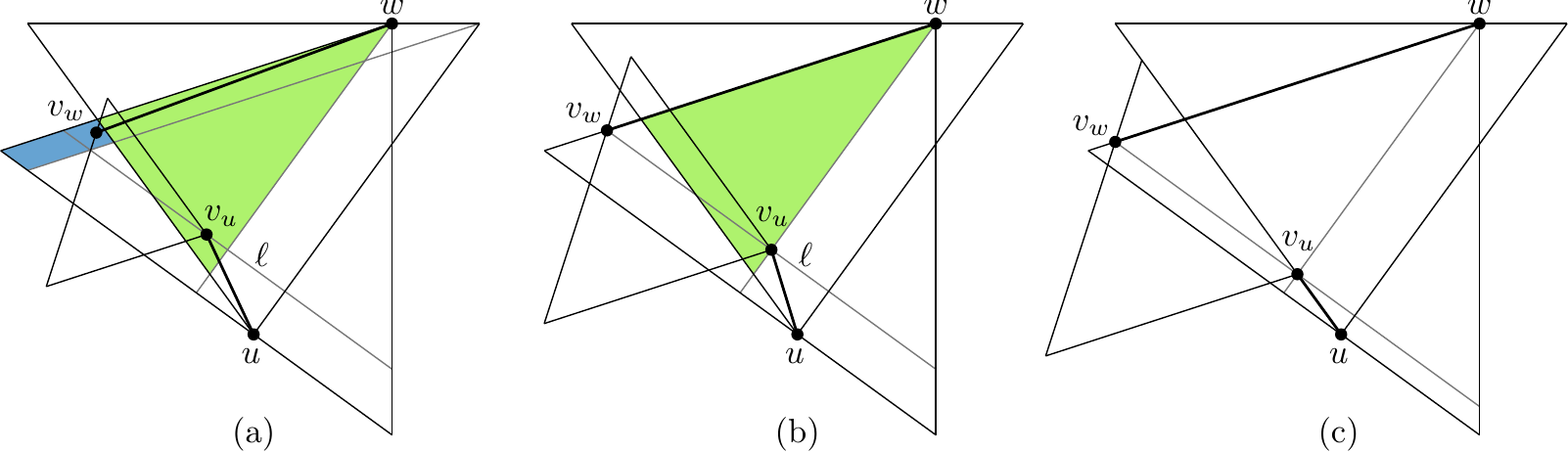}
  \caption{(a) The regions where $v_u$ (light) and $v_w$ (dark) can lie. (b) The worst case when $v_u$ lies on a given line $\ell$. (c) The worst case for a fixed position of $w$.}
  \label{fig:complex}
 \end{figure}

 \paragraph{Case 4d.} $w$ lies in $C_1^{v_u}$ and $v_u$ lies in $C_3^w$. We would like to apply our inductive hypothesis to $\T{v_u}{v_w}$, resulting in a path between $v_u$ and $v_w$ of length at most $c \cdot |\T{v_u}{v_w}|$. The edges $(w, v_w)$ and $(u, v_u)$ complete this to a path between $u$ and $w$, giving a total length of at most $|uv_u| + c \cdot |\T{v_u}{v_w}| + |v_ww|$.

 First, note that $v_u$ cannot lie in $\T{w}{v_w}$, as this region is empty by definition. Since $v_w$ lies in $C_4^u$, this means that $v_w$ must lie in $C_4^{v_u}$. We first show that $\T{v_u}{v_w}$ is always smaller than $\T{u}{w}$, which means that we are allowed to use induction. Given any initial position for $v_u$, consider the line $\ell$ through $v_u$, perpendicular to the bisector of $C_3$ (see~Figure~\ref{fig:complex}a). Since $v_w$ cannot be further from $w$ than $v_u$, the size of $\T{v_u}{v_w}$ is maximized when $v_w$ lies on the intersection of $\ell$ and the top boundary of $\T{w}{u}$. We can increase $|\T{v_u}{v_w}|$ further by moving $v_u$ along $\ell$ until it reaches the bisector of $C_3^w$ (see~Figure~\ref{fig:complex}b). Since the top boundary of $\T{w}{u}$ and the bisector of $C_3^w$ approach each other as they get closer to $w$, the size of $\T{v_u}{v_w}$ is maximized when $v_u$ lies on the bottom boundary of $\T{w}{u}$ (ignoring for now that this would move $v_u$ out of $\T{u}{w}$). Now it is clear that $|\T{v_u}{v_w}| < |\T{u}{v_w}|$. Since we already established that $\T{u}{v_w}$ is smaller than $\T{u}{w}$ in the proof of Theorem~\ref{thm:connected}, this holds for $\T{v_u}{v_w}$ as well and we can use induction.

 All that is left is to bound the total length of the path. Given any initial position of $v_u$, the path length is maximized when we place $v_w$ at the intersection of $\ell$ and the top boundary of $\T{w}{u}$, as this maximizes both $|\T{v_u}{v_w}|$ and $|wv_w|$. When we move $v_u$ away from $v_w$ along $\ell$ by $\Delta$, $|uv_u|$ decreases by at most $\Delta$, while $|\T{v_u}{v_w}|$ increases by $\sin \frac{3\pi}{5} / \sin \frac{3\pi}{10} \cdot \Delta > \Delta$. Since $c > 1$, this increases the total path length. Therefore the worst case again occurs when $v_u$ lies on the bisector of $C_3^w$, as depicted in Figure~\ref{fig:complex}b. Moving $v_u$ down along the bisector of $\T{w}{u}$ by $\Delta$ decreases $|uv_u|$ by at most $\Delta$, while increasing $|wv_w|$ by $1 / \sin \frac{3\pi}{10} \cdot \Delta > \Delta$ and increasing $|\T{v_u}{v_w}|$. Therefore this increases the total path length and the worst case occurs when $v_u$ lies on the left boundary of $\T{u}{w}$ (see~Figure~\ref{fig:complex}c).

 Finally, consider what happens when we move $v_u$ $\Delta$ towards $u$, while moving $w$ and $v_w$ such that the construction stays intact. This causes $w$ to move to the right. Since $v_u$, $w$ and the left corner of $\T{u}{w}$ form an isosceles triangle with apex $v_u$, this also moves $v_u$ $\Delta$ further from $w$. We saw before that moving $v_u$ away from $w$ increases the size of $\T{v_u}{v_w}$. Finally, it also increases $|wv_w|$ by $1 / \sin \frac{3\pi}{10} \cdot \Delta > \Delta$. Thus, the increase in $|wv_w|$ cancels the decrease in $|uv_u|$ and the total path length increases. Therefore the worst case occurs when $v_u$ lies almost on $u$ and $v_w$ lies in the corner of $\T{w}{u}$, which is symmetric to the worst case of Case~1. Thus the theorem holds by the same analysis.

 \paragraph{Case 4e.} $w$ lies in $C_1^{v_u}$ and $v_u$ lies in $C_4^w$. We split this case into three final subcases, based on the position of $v_u$. These cases are illustrated in Figure~\ref{fig:complex2}a. Note that $v_u$ cannot lie in $C_2$ or $C_3$ of $v_w$, as it lies above $v_w$. It also cannot lie in $C_4^{v_w}$, as $C_4^{v_w}$ is completely contained in $C_4^u$, whereas $v_u$ lies in $C_0^u$. Thus the cases presented below are exhaustive.

 \paragraph{Case 4e-1.} $|\T{w}{v_u}| \leq \frac{c - 1}{c} \cdot |\T{u}{w}|$. If $\T{w}{v_u}$ is small enough, we can apply our inductive hypothesis to obtain a path between $v_u$ and $w$ of length at most $c \cdot |\T{w}{v_u}|$. Since there is a direct edge between $u$ and $v_u$, we obtain a path between $u$ and $w$ of length at most $|uv_u| + c \cdot |\T{w}{v_u}|$. Any edge from $u$ to a point inside $\T{u}{w}$ has length at most $|\T{u}{w}|$, so we can bound the length of the path as follows.
 \[
  |uv_u| + c \cdot |\T{w}{v_u}|
  ~~\leq~~ |\T{u}{w}| + c \cdot \frac{c - 1}{c} \cdot |\T{u}{w}|
  ~~=~~ |\T{u}{w}| + (c - 1) \cdot |\T{u}{w}|
  ~~=~~ c \cdot |\T{u}{w}|
 \]

 \bigskip
 In the other two cases, we use induction on $\T{v_w}{v_u}$ to obtain a path between $v_w$ and $v_u$ of length at most $c \cdot |\T{v_w}{v_u}|$. The edges $(u, v_u)$ and $(w, v_w)$ complete this to a (self-intersecting) path between $u$ and $w$. We can bound the length of these edges by the size of the canonical triangle that contains them, as follows.
 \[
  |uv_u| + |wv_w|
  ~~\leq~~ |\T{u}{w}| + |\T{w}{u}|
  ~~\leq~~ |\T{u}{w}| + \frac{1}{\cos \frac{\pi}{5}} \cdot |\T{u}{w}|
  ~~=~~ \sqrt{5} \cdot |\T{u}{w}|
 \]
 All that is left now is to bound the size of $\T{v_w}{v_u}$ and express it in terms of $\T{u}{w}$.

 \begin{figure}[ht]
  \centering
  \includegraphics{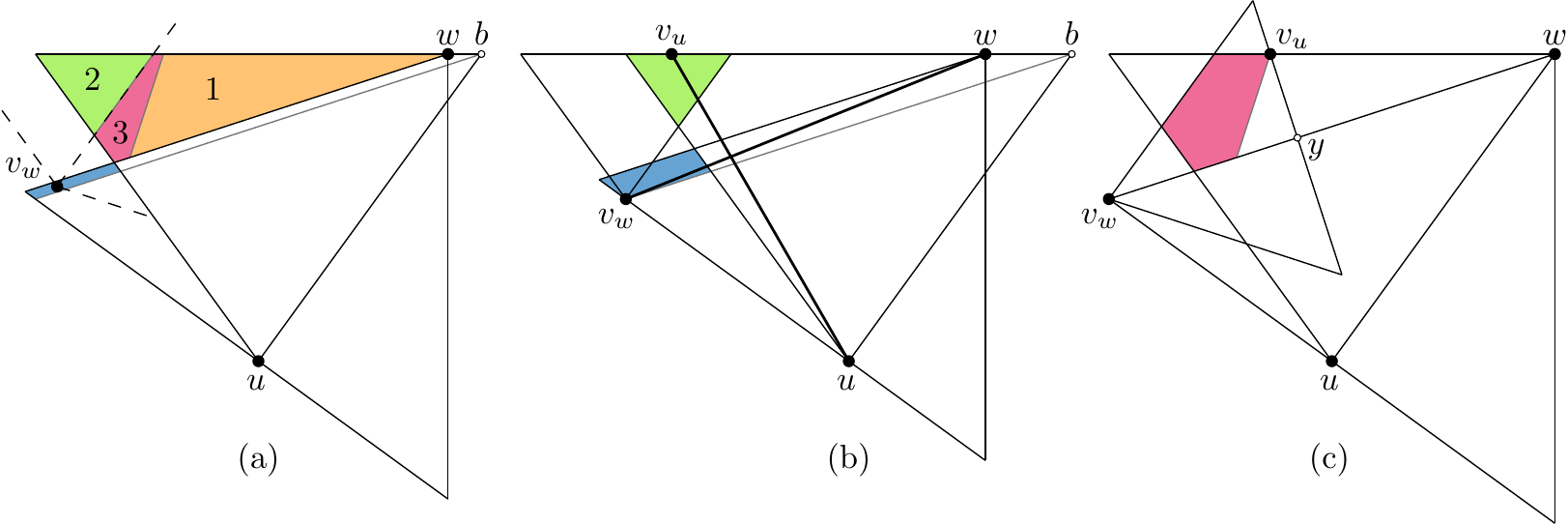}
  \caption{(a) The three subcases for the position of $v_u$. (b) The situation that maximizes $\T{v_w}{v_u}$ when $v_u$ lies in $C_0^{v_w}$. (c) The worst case when $v_u$ lies in $C_1^{v_w}$.}
  \label{fig:complex2}
 \end{figure}

 \paragraph{Case 4e-2.} $v_u$ lies in $C_0^{v_w}$. In this case, the size of $\T{v_w}{v_u}$ is maximal when $v_u$ lies on the top boundary of $\T{u}{w}$ and $v_w$ lies at the lowest point in its possible region: the left corner of $\T{b}{u}$ (see~Figure~\ref{fig:complex2}b). Now we can express $|\T{v_w}{v_u}|$ as follows.
 \[
  |\T{v_w}{v_u}|
  ~~=~~ \frac{\sin \frac{\pi}{10}}{\sin \frac{7\pi}{10}} \cdot |bv_w|
  ~~=~~ \frac{\sin \frac{\pi}{10}}{\sin \frac{7\pi}{10}} \cdot \frac{1}{\cos \frac{\pi}{5}} \cdot |\T{u}{w}|
  ~~=~~ 2 \left( \sqrt{5} - 2 \right) \cdot |\T{u}{w}|
 \]
 Since $2 \left( \sqrt{5} - 2 \right) < 1$, we can use induction. The total path length is bounded by $c \cdot |\T{u}{w}|$ for
 \[
  c
  ~~\geq~~ \frac{\sqrt{5}}{1 - 2 \left( \sqrt{5} - 2 \right)}
  ~~=~~ 2 + \sqrt{5}
  ~~\approx~~ 4.236.
 \]

 \paragraph{Case 4e-3.} $v_u$ lies in $C_1^{v_w}$. Since $|\T{w}{v_u}| > \frac{c - 1}{c} \cdot |\T{u}{w}|$, $\T{v_w}{v_u}$ is maximal when $v_w$ lies on the left corner of $\T{w}{u}$ and $v_u$ lies on the top boundary of $\T{u}{w}$, such that $|\T{w}{v_u}| = \frac{c - 1}{c} \cdot |\T{u}{w}|$ (see~Figure~\ref{fig:complex2}c). Let $y$ be the intersection of $\T{v_w}{v_u}$ and $\T{w}{u}$. Note that since $v_w$ lies on the corner of $\T{w}{u}$, $y$ is also the midpoint of the side of $\T{v_w}{v_u}$ opposite $v_w$. We can express the size of $\T{v_w}{v_u}$ as follows.
 \[
  |\T{v_w}{v_u}|
  ~~=~~\frac{|v_wy|}{\cos \frac{\pi}{5}}
  ~~=~~\frac{|wv_w| - |wy|}{\cos \frac{\pi}{5}}
  ~~=~~\frac{\displaystyle\frac{|\T{u}{w}|}{\cos \frac{\pi}{5}} - \cos {\textstyle \frac{\pi}{10}} \cdot |wv_u|}{\cos \frac{\pi}{5}}
 \]
 \[
  =~~\frac{\displaystyle\frac{|\T{u}{w}|}{\cos \frac{\pi}{5}} - \cos {\textstyle \frac{\pi}{10}} \cdot \frac{\sin \frac{3\pi}{10}}{\sin \frac{3\pi}{5}} \cdot |\T{w}{v_u}|}{\cos \frac{\pi}{5}}
  ~~=~~\frac{\displaystyle\frac{|\T{u}{w}|}{\cos \frac{\pi}{5}} - \cos {\textstyle \frac{\pi}{10}} \cdot \frac{\sin \frac{3\pi}{10}}{\sin \frac{3\pi}{5}} \cdot \frac{c - 1}{c} \cdot |\T{u}{w}|}{\cos \frac{\pi}{5}}
 \]
 \[
  =~~\left( \frac{1}{c} + 5 - 2 \sqrt{5} \right) \cdot |\T{u}{w}|
 \]
 Thus we can use induction for $c > 1 / \left(2 \sqrt{5} - 4\right) \approx 2.118$ and the total path length can be bounded by $c \cdot |\T{u}{w}|$ for
 \[
  c
  ~~\geq~~\frac{\sqrt{5} + 1}{2 \sqrt{5} - 4}
  ~~=~~ \frac{1}{2}\left( 7 + 3 \sqrt{5} \right)
  ~~\approx~~ 6.854.
 \]
\end{myproof}

\noindent Using this result, we can compute the exact spanning ratio.

\begin{theorem}
 \label{thm:spanner}
 The \graph is a spanner with spanning ratio at most \valsr.
\end{theorem}
\begin{myproof}
 Given two vertices $u$ and $w$, we know from Lemma~\ref{lem:spanningPath} that there is a path between them of length at most $c \cdot \min\left(|\T{u}{w}|, |\T{w}{u}|\right)$, where $c = \valc$. This gives an upper bound on the spanning ratio of $c \cdot \min\left(|\T{u}{w}|, |\T{w}{u}|\right) / |uw|$. We assume without loss of generality that $w$ lies in the right half of $C_0^u$. Let $\alpha$ be the angle between the bisector of $C_0^u$ and the line $uw$ (see~Figure~\ref{fig:canon}b). In the proof of Theorem~\ref{thm:connected}, we saw that we can express $|\T{w}{u}|$ and $|uw|$ in terms of $\alpha$ and $|\T{u}{w}|$, as $|\T{w}{u}| = (\cos (\frac{\pi}{5} - \alpha) / \cos \alpha) \cdot |\T{u}{w}|$ and $|uw| = (\cos \frac{\pi}{5} / \cos \alpha) \cdot |\T{u}{w}|$, respectively. Using these expressions, we can write the spanning ratio in terms of $\alpha$.
 \[
  \frac{c \cdot \min\left(|\T{u}{w}|, |\T{w}{u}|\right)}{|uw|}
  ~~=~~ \frac{c \cdot \min\left(|\T{u}{w}|, \frac{\cos \left( \frac{\pi}{5} - \alpha \right)}{\cos \alpha} \cdot |\T{u}{w}|\right)}{\frac{\cos \frac{\pi}{5}}{\cos \alpha} \cdot |\T{u}{w}|}
  ~~=~~ \frac{c}{\cos \frac{\pi}{5}} \cdot \min\left(\cos \alpha, \cos \left( {\textstyle \frac{\pi}{5}} - \alpha \right)\right)
 \]
 To get an upper bound on the spanning ratio, we need to maximize the minimum of $\cos \alpha$ and $\cos \left(\frac{\pi}{5} - \alpha\right)$. Since for $\alpha \in [0, \pi / 5]$, one is increasing and the other is decreasing, this maximum occurs at $\alpha = \pi/10$, where they are equal. Thus, our upper bound becomes
 \[
  \frac{c}{\cos \frac{\pi}{5}} \cdot \cos {\textstyle \frac{\pi}{10}}
  ~~=~~\sqrt{50 + 22 \sqrt{5}}
  ~~\approx~~9.960.
 \]
 \vspace{-2em}
\end{myproof}

\section{Lower bound}

\begin{wrapfigure}{r}{0.4\textwidth}
  \vspace{-2em}
  \begin{center}
    \includegraphics{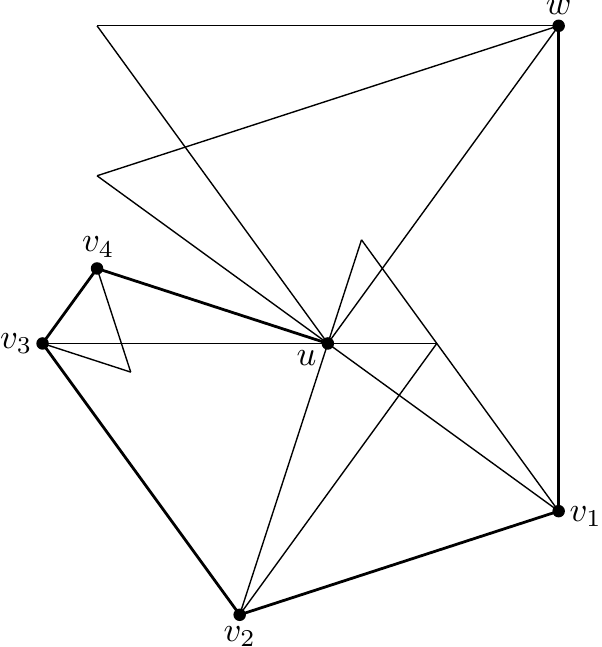}
  \end{center}
  \vspace{-1em}
  \caption{A path with a large spanning ratio.}
  \label{fig:lbpath}
\end{wrapfigure}

In this section, we derive a lower bound on the spanning ratio of the \graph.

\begin{theorem}
 \label{thm:lowerbound}
 The \graph has spanning ratio at least $\vallb$.
\end{theorem}
\begin{myproof}
 For the lower bound, we present and analyze a path between two vertices that has a large spanning ratio. The path has the following structure (illustrated in Figure~\ref{fig:lbpath}).

 The path can be thought of as being directed from $w$ to $u$. First, we place $w$ in the right corner of $\T{u}{w}$. Then we add a vertex $v_1$ in the bottom corner of $\T{w}{u}$. We repeat this two more times, each time adding a new vertex in the corner of $\T{v_i}{u}$ furthest away from $u$. The final vertex $v_4$ is placed on the top boundary of $C_1^{v_3}$, such that $u$ lies in $C_1^{v_4}$. Since we know all the angles involved, we can compute the length of each edge, taking $|uw| = 1$ as baseline.
 \[
  |wv_1| = \frac{1}{\cos \frac{\pi}{5}}~~~~~|v_1v_2| = |v_2v_3| = 2 \sin {\textstyle \frac{\pi}{5}} \tan {\textstyle \frac{\pi}{5}}~~~~~|v_3v_4| = \frac{\sin \frac{\pi}{10}}{\sin \frac{3\pi}{5}} \tan {\textstyle \frac{\pi}{5}}~~~~~|v_4u| = \frac{\sin \frac{3\pi}{10}}{\sin \frac{3\pi}{5}} \tan {\textstyle \frac{\pi}{5}}
 \]
 Since we set $|uw| = 1$, the spanning ratio is simply $|wv_1| + |v_1v_2| + |v_2v_3| + |v_3v_4| + |v_4u| = \textstyle\frac{1}{2}(11\sqrt{5} -17) \approx 3.798$. Note that the \graph with just these 5 vertices would have a far smaller spanning ratio, as there would be a lot of shortcut edges. However, a graph where this path is the shortest path between two vertices can be found in Appendix~\ref{app:lbgraph}.
\end{myproof}

\section{Lower bound on $\theta$-routing}
\label{sec:lbrouting}

\input{thetarouting}

\section{Conclusions}
\label{sec:Conclusions}

We showed that there is a path between every pair of vertices in the \graph, of length at most \valsr times the straight-line distance between them. This is the first constant upper bound on the spanning ratio of the \graph, proving that it is a geometric spanner. We also presented a \graph with spanning ratio arbitrarily close to \vallb, thereby giving a lower bound on the spanning ratio. There is still a significant gap between these bounds, which is caused by the upper bound proof mostly ignoring the main obstacle to improving the lower bound: that every edge requires at least one of its canonical triangles to be empty. Hence we believe that the true spanning ratio is closer to the lower bound.

While our proof for the upper bound on the spanning ratio returns a spanning path between the two vertices, it requires knowledge of the neighbours of both the current vertex and the destination vertex. This means that the proof does not lead to a local routing strategy that can be applied in, say, a wireless setting. This raises the question whether it is possible to route \emph{competitively} on this graph, i.e. to discover a spanning path from one vertex to another by using only information local to the current vertices visited so far.

\bibliographystyle{unsrt}
\bibliography{theta5}

\newpage
\appendix

\section{Lower bound on the spanning ratio}
\label{app:lbgraph}

\begin{tabular}{|c|p{0.732\textwidth}|c|}
\hline
\# & Action & Shortest path \\
\hline
1 & Start with a vertex $v_{1}$. & - \\
2 & Add $v_{2}$ in $C_0^u$, such that $v_{2}$ is arbitrarily close to the top right corner of $\T{v_{1}}{v_{2}}$. & $v_{1}v_{2}$\\
3 & Remove edge $(v_{1}, v_{2})$ by adding two vertices, $v_{3}$ and $v_{4}$, arbitrarily close to the counter-clockwise corners of $\T{v_{1}}{v_{2}}$ and $\T{v_{2}}{v_{1}}$. & $v_{1}v_{4}v_{2}$\\
4 & Remove edge $(v_{1}, v_{4})$ by adding two vertices, $v_{5}$ and $v_{6}$, arbitrarily close to the clockwise corner of $\T{v_{1}}{v_{4}}$ and the counter-clockwise corner of $\T{v_{4}}{v_{1}}$. & $v_{1}v_{3}v_{2}$\\
5 & Remove edge $(v_{2}, v_{3})$ by adding two vertices, $v_{7}$ and $v_{8}$, arbitrarily close to the clockwise corner of $\T{v_{2}}{v_{3}}$ and the counter-clockwise corner of $\T{v_{3}}{v_{2}}$. & $v_{1}v_{6}v_{4}v_{2}$\\
6 & Remove edge $(v_{1}, v_{6})$ by adding two vertices, $v_{9}$ and $v_{10}$, arbitrarily close to the clockwise corner of $\T{v_{1}}{v_{6}}$ and the counter-clockwise corner of $\T{v_{6}}{v_{1}}$. & $v_{1}v_{5}v_{4}v_{2}$\\
7 & Remove edge $(v_{4}, v_{5})$ by adding two vertices, $v_{11}$ and $v_{12}$, arbitrarily close to the counter-clockwise corner of $\T{v_{4}}{v_{5}}$ and the clockwise corner of $\T{v_{5}}{v_{4}}$. & $v_{1}v_{5}v_{6}v_{4}v_{2}$\\
8 & Remove edge $(v_{5}, v_{6})$ by adding two vertices, $v_{13}$ and $v_{14}$, arbitrarily close to the counter-clockwise corner of $\T{v_{5}}{v_{6}}$ and the clockwise corner of $\T{v_{6}}{v_{5}}$. & $v_{1}v_{5}v_{14}v_{6}v_{4}v_{2}$\\
9 & Remove edge $(v_{5}, v_{14})$ by adding two vertices, $v_{15}$ and $v_{16}$, arbitrarily close to the counter-clockwise corner of $\T{v_{5}}{v_{14}}$ and the clockwise corner of $\T{v_{14}}{v_{5}}$. & $v_{1}v_{5}v_{13}v_{6}v_{4}v_{2}$\\
10 & Remove edge $(v_{6}, v_{13})$ by adding two vertices, $v_{17}$ and $v_{18}$, arbitrarily close to the clockwise corner of $\T{v_{6}}{v_{13}}$ and the counter-clockwise corner of $\T{v_{13}}{v_{6}}$. & $v_{1}v_{3}v_{8}v_{2}$\\
11 & Remove edge $(v_{2}, v_{8})$ by adding a vertex $v_{19}$ in the union of, and arbitrarily close to the intersection point of $\T{v_{2}}{v_{8}}$ and $\T{v_{8}}{v_{2}}$. & $v_{1}v_{3}v_{7}v_{2}$\\
12 & Remove edge $(v_{3}, v_{7})$ by adding two vertices, $v_{20}$ and $v_{21}$, arbitrarily close to the counter-clockwise corner of $\T{v_{3}}{v_{7}}$ and the clockwise corner of $\T{v_{7}}{v_{3}}$. & $v_{1}v_{5}v_{12}v_{2}$\\
13 & Remove edge $(v_{2}, v_{12})$ by adding a vertex $v_{22}$ arbitrarily close to the counter-clockwise corner of $\T{v_{2}}{v_{12}}$. & $v_{1}v_{10}v_{6}v_{4}v_{2}$\\
14 & Remove edge $(v_{1}, v_{10})$ by adding a vertex $v_{23}$ in the union of $\T{v_{1}}{v_{10}}$ and $\T{v_{10}}{v_{1}}$, arbitrarily close to the top boundary of $C_1^{v_{10}}$, and such that $v_{1}$ lies in $C_1^{v_{23}}$, arbitrarily close to the bottom boundary. & $v_{1}v_{5}v_{12}v_{4}v_{2}$\\
15 & Remove edge $(v_{4}, v_{12})$ by adding two vertices, $v_{24}$ and $v_{25}$, arbitrarily close to the counter-clockwise corner of $\T{v_{4}}{v_{12}}$ and the clockwise corner of $\T{v_{12}}{v_{4}}$. & $v_{1}v_{5}v_{13}v_{14}v_{6}v_{4}v_{2}$\\
16 & Remove edge $(v_{13}, v_{14})$ by adding two vertices, $v_{26}$ and $v_{27}$, arbitrarily close to the clockwise corner of $\T{v_{13}}{v_{14}}$ and the counter-clockwise corner of $\T{v_{14}}{v_{13}}$. & $v_{1}v_{9}v_{18}v_{6}v_{4}v_{2}$\\
17 & Remove edge $(v_{9}, v_{18})$ by adding two vertices, $v_{28}$ and $v_{29}$, arbitrarily close to the clockwise corner of $\T{v_{9}}{v_{18}}$ and the counter-clockwise corner of $\T{v_{18}}{v_{9}}$. & $v_{1}v_{5}v_{16}v_{11}v_{4}v_{2}$\\
18 & Remove edge $(v_{11}, v_{16})$ by adding two vertices, $v_{30}$ and $v_{31}$, arbitrarily close to the counter-clockwise corner of $\T{v_{11}}{v_{16}}$ and the clockwise corner of $\T{v_{16}}{v_{11}}$. & $v_{1}v_{23}v_{10}v_{6}v_{4}v_{2}$\\
\hline
\end{tabular}

 \begin{figure}[ht]
  \centering
  \includegraphics{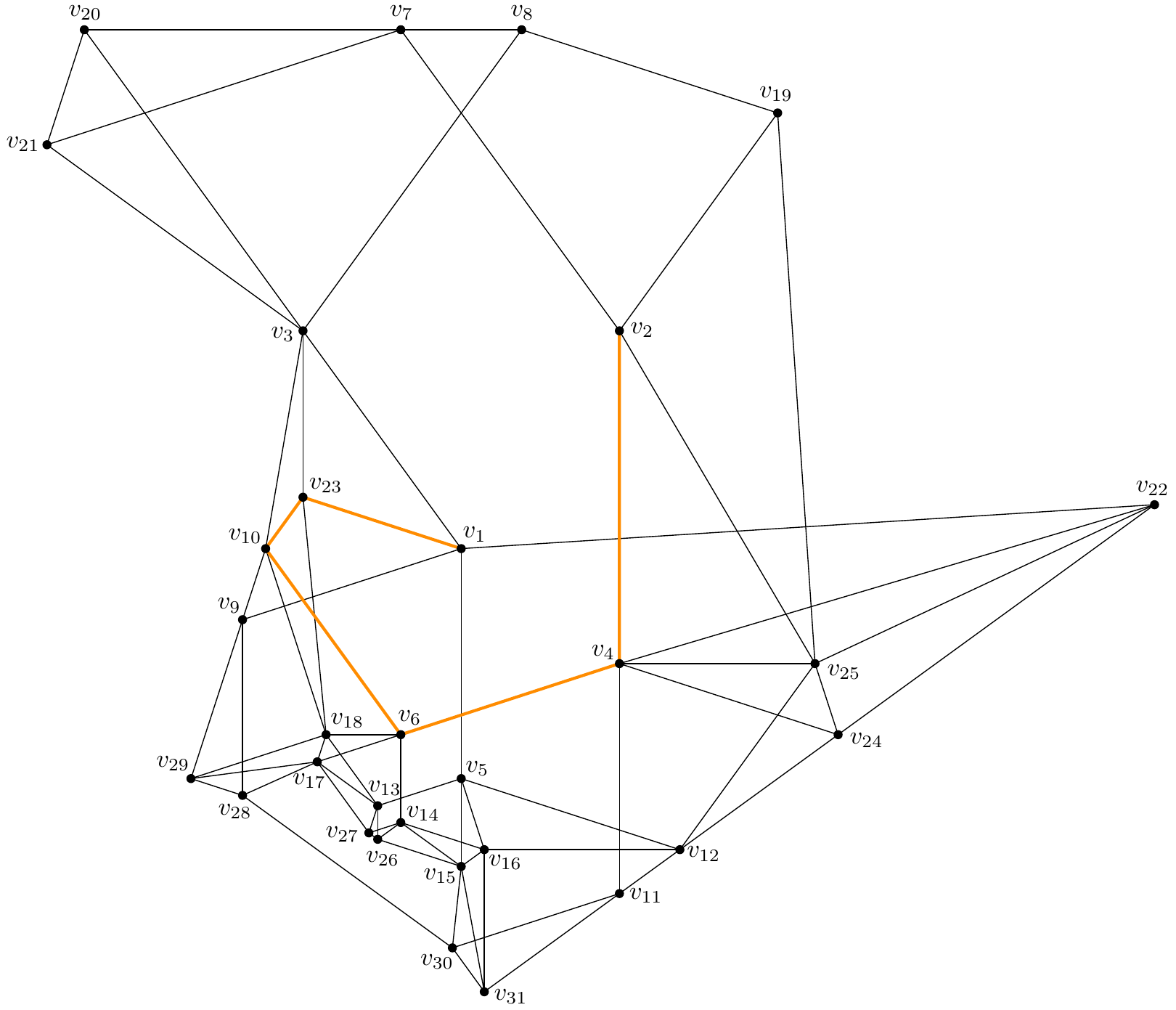}
  \caption{A \graph with a spanning ratio that matches the lower bound. The shortest path between $v_1$ and $v_2$ is indicated in orange.}
  \label{fig:lb}
 \end{figure}

\end{document}

%% file: thetarouting.tex
In this section, we show that always following the edge to the closest vertex in the cone that contains the destination can generate very long paths in the \graph. More formally, we look at the \emph{competitiveness} of this routing algorithm. A routing algorithm is $c$-competitive on a graph $G$ if for each pair of vertices in the graph, the routing algorithm finds a path of length at most $c$ times the Euclidean distance between the two vertices. 

We look at the competitiveness of the \emph{$\theta$-routing} algorithm, the standard routing algorithm for $\theta$-graphs with at least seven cones: From the current vertex $u$, follow the edge to the closest vertex in \T{u}{w}, where $w$ is the destination. This step is repeated until the destination is reached. We construct a \graph for which the $\theta$-routing algorithm returns a path with spanning ratio $\Omega(n)$.

\begin{lemma}
  The $\theta$-routing algorithm is not $o(n)$-competitive on the \graph. 
\end{lemma}
\begin{myproof}
  We construct the lower bound example on the competitiveness of the $\theta$-routing algorithm on the \graph by repeatedly extending the routing path from source $u$ to destination $w$. First, we place $w$ such that the angle between $u w$ and the bisector of \T{u}{w} is $\theta/4$. To ensure that the $\theta$-routing algorithm does not follow the edge between $u$ and $w$, we place a vertex $v_1$ in the upper corner of \T{u}{w} that is furthest from $w$. Next, to ensure that the $\theta$-routing algorithm does not follow the edge between $v_1$ and $w$, we place a vertex $v_1'$ in the upper corner of \T{v_1}{w} that is furthest from $w$. We repeat this step until we have created a cycle around $w$ (see Figure~\ref{fig:thetaRouting}a). 

  \begin{figure}[ht]
    \centering
    \includegraphics{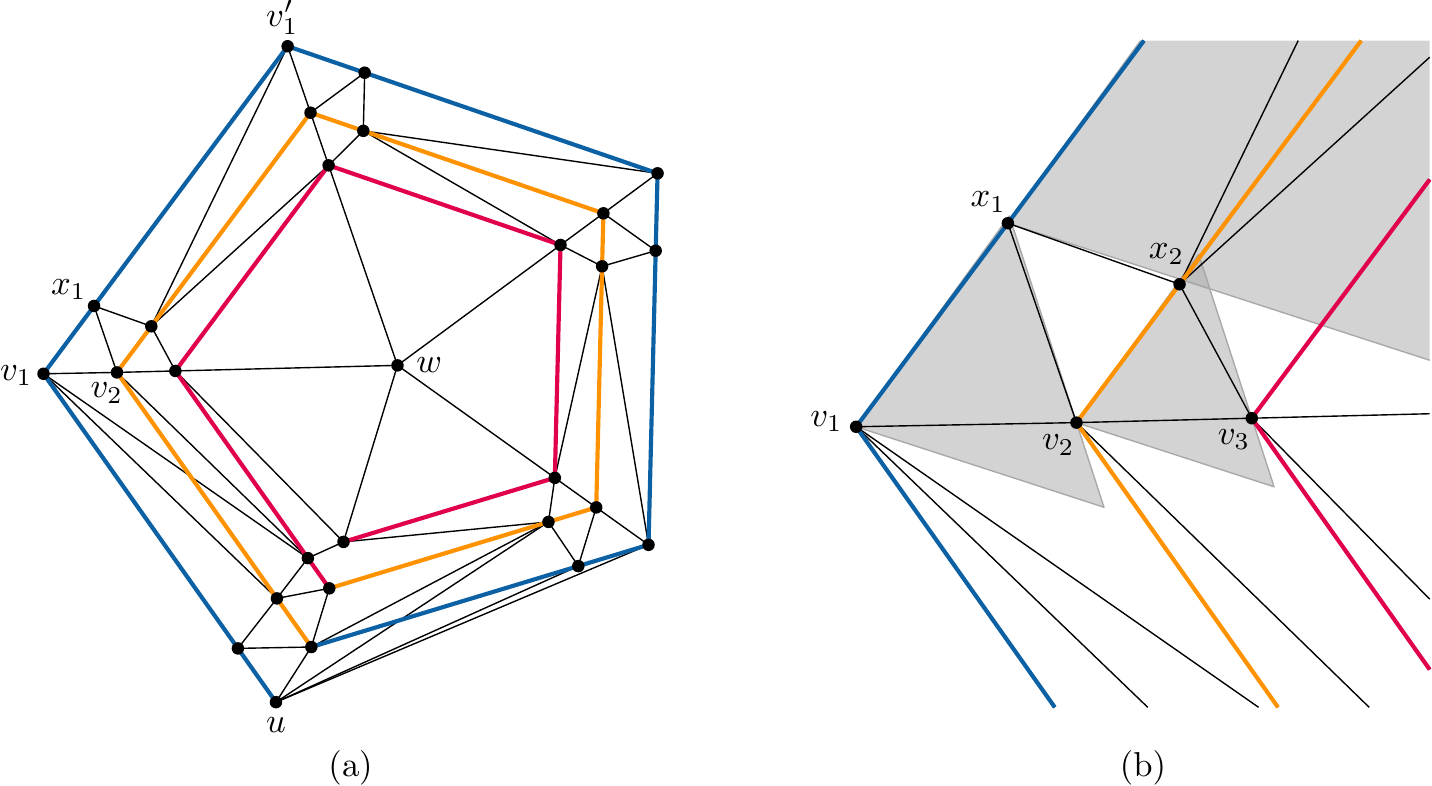}
    \caption{(a) A lower bound example for $\theta$-routing on the \graph, consisting of three cycles: the first cycle is coloured blue, the second cycle is coloured orange, and the third cycle is coloured red. (b) The placement of vertices such that previous cycles stay intact when adding a new cycle.}
    \label{fig:thetaRouting}
  \end{figure}

  To extend the routing path further, we again place a vertex $v_2$ in the corner of the current canonical triangle. To ensure that the routing algorithm still routes to $v_1$ from $u$, we place $v_2$ slightly outside of \T{u}{v_1}. However, another problem arises: vertex $v_1'$ is no longer the vertex closest to $v_1$ in \T{v_1}{w}, as $v_2$ is closer. To solve this problem, we also place a vertex $x_1$ in \T{v_1}{v_2} such that $v_1'$ lies in \T{x_1}{w} (see Figure~\ref{fig:thetaRouting}b). By repeating this process four times, we create a second cycle around $w$.

  To add more cycles around $w$, we repeat the same process as described above: place a vertex in the corner of the current canonical triangle and place an auxiliary vertex to ensure that the previous cycle stays intact. Note that when placing $x_i$, we also need to ensure that it does not lie in \T{x_{i-1}}{w}, to prevent shortcuts from being formed (see Figure~\ref{fig:thetaRouting}b). This means that in general $x_i$ does not lie arbitrarily close to the corner of \T{v_i}{v_{i+1}}. 

  This way we need to add auxiliary vertices only to the $(k-1)$-th cycle, when adding the $k$-th cycle, hence we can add an additional cycle using only a constant number of vertices. Since we can place the vertices arbitrarily close to the corners of the canonical triangles, we ensure that the distance to $w$ stays almost the same, regardless of the number of cycles. Hence, each `step' of the form $v_i$ to $v_i'$ via $x_i$ has length $\cos(\theta/4) / \cos(\theta/2) \cdot |v_i w|$. Since $\cos(\theta/4) / \cos(\theta/2)$ is greater than 1 for the \graph and $|v_i w|$ can be arbitrarily close to $|u w|$, every step has length greater than $|u w|$. Using $n$ points, we can construct $n/2$ of these steps and the total length of the path followed by the $\theta$-routing algorithm is greater than $n/2 \cdot |u w|$. Thus the $\theta$-routing algorithm is not $o(n)$-competitive on the constructed graph.
\end{myproof}